\documentclass[12pt,titlepage]{article}
\usepackage[english]{babel}
\usepackage[dvips]{graphicx}

\newcommand{\eq}{{\, \equiv\, }}
\newcommand{\fr}[1]{
             \frac{#1}}
\newcommand{\bea}{\begin{eqnarray}}
\newcommand{\eea}{\end{eqnarray}}

\newcommand{\chibar}{\overline{\chi}}

\newcommand{\ket}{{\rangle }}

\newcommand{\bra}{{\langle }}
\newcommand{\gc}{\bra\fr{\alpha_s}{\pi}G^2\ket}

\newcommand{\ga}{{g_{{\mathcal A}}}}

\begin{document}

\begin{titlepage}

\begin{center}

{\Large \bf  Color suppressed contributions to the decay modes\\
\vspace{0.3cm}
 $B_{d,s} \to D_{s,d} D_{s,d} \; ,$  $B_{d,s} \to D_{s,d} D^*_{s,d} \; ,$  
and $B_{d,s} \to D^*_{s,d} D^*_{s,d} \; . $}
  \\
\vspace{1cm}

{\large \bf J. O. Eeg$^{a}$, S. Fajfer$^{b,c}$,  A. Prapotnik$^{c}$\\}
\vspace{.5cm}

{\it a) Dept. of Physics, Univ. of Oslo, P.O. Box 1048
Blindern, N-0316 Oslo, Norway\\}
\vspace{.5cm}

{\it b)  Department of Physics, University of Ljubljana, Jadranska 19,
1000 Ljubljana, Slovenia\\}
\vspace{.5cm}

{\it c) J. Stefan Institute, Jamova 39, P. O. Box 3000, 1001 Ljubljana, Slovenia\\}
\vspace{.5cm}

\end{center}

\vspace{0.25cm}

\centerline{\large \bf ABSTRACT}

\vspace{0.2cm}


The amplitudes for decays of the type 
$B_{d,s} \to D_{s,d} D_{s,d}$,
have no factorizable contributions,
while $B_{d,s} \to D_{s,d} D^*_{s,d}$,  
and $B_{d,s} \to D^*_{s,d} D^*_{s,d}$
have relatively small factorizable contributions through the
annihilation mechanism.
The dominant contributions to the decay amplitudes 
arise from  chiral loop contributions and 
tree level amplitudes which can be obtained in terms of  soft gluon emissions
forming a gluon condensate.
We predict that the 
branching ratios   for the processes $\bar B^0_d \to D_s^+ D_s^-$, 
  $\bar B^0_d \to D_s^{+*} D_s^- $ and 
 $\bar B^0_d \to D_s^+ D_s^{-*}$ are all of order $(2- 3) \times 10^{-4}$,
while $\bar B^0_s \to D_d^+ D_d^-$,
  $\bar B^0_s \to D_d^{+*} D_d^- $ and 
 $\bar B^0_s \to D_d^+ D_d^{-*}$ are  of order $(4- 7) \times 10^{-3}$.
We obtain branching ratios for two $D^*$'s in the final state  of order
two times bigger.


\end{titlepage}  

\section{Introduction}

Recently, many theoretical and experimental studies in
$B$-meson physics have been done.  The most extensive investigations have been 
 done for cases of
$B$-decay modes into the light mesons (e.g. $B \to \pi \pi$, $B \to K
\pi$) due to the their importance in the 
determination of Cabibbo-Kobayashi-Maskawa (CKM) matrix
elements. 
On the other hand $B$-decays into charmed mesons also  present
an important issue for experimental and theoretical analysis.
 As it was pointed out by the authors of
\cite{PhaCo,FSI}, final state interactions (FSI) including  charmed
intermediate states can give significant contributions to the decay amplitudes, 
especially in
the case of the $B \to K \pi$ decay modes \cite{PhaCo}. 
In addition, branchig ratios for $B_d$
decay into $D^- D_s^+$, $D^{-*} D_s^+$, $D^- D_s^{*+}$ and $D^{*-}
D_s^{*+}$ \cite{PDG} have been  measured. A previous theoretical study of 
the decay mode  $B_d \to D^- D_s^+$ \cite{Fleischer} showed that the non-factorizable
contributions coming from chiral loops give an increase of order 10\%  in
the rate compared to the factorized limit. The rate found in 
 \cite{Fleischer} is in
good agreement with the experimental result.
On the other hand, the $\bar B^0 \to D^+_s D^-_s$ and $\bar B^0_s \to
D^+ D^-$ decay modes have no factorizable amplitudes and they are
realized only through non-factorizable contributions as it was shown
in \cite{EFH}.  At the quark level, these decays a priori proceed through
the annihilation mechanism $b \bar{s} \rightarrow c
\bar{c}$ and $b \bar{d} \rightarrow c \bar{c}$, respectively. 
Within the factorized limit this mechanism will give a zero amplitude
due to current conservation, as in the case of  $D^0 \rightarrow K^0
\overline{K^0}$ \cite{EFZ}. Therefore $B_{d,s} \to D_{s,d} D_{s,d} \, $,
$B_{d,s} \to D_{s,d} D^*_{s,d} \, $, and $B_{d,s} \to D^*_{s,d} D^*_{s,d}$
present a fertile ground for the investigation of  non-factorizable $1/N_c$
suppressed contributions.

In this paper we want to extend our previous study of 
$B_{d,s} \to D_{s,d} D_{s,d}$ 
decays to cases with one or two $D^*$'s in the final state.  Namely,
at $B$-factories the decays to one pseudoscalar and one vector
$D$-meson
 are easier accessable  due to better statistical accuracy (they can
be reconstructed more inclusively than decays to two pseudoscalars)
\cite{GoKr}.
In this case non-factorizable
contributions arise  due to the annihilation mechanism.
 However, its amplitude is suppressed by a 
numerically less favorable combination of  Wilson coefficients and is 
expected to be  of the same order of magnitude 
as non-factorizable contributions.
Since the  energy release for $B$-decays to two charm mesons is 
relatively small (of order  $1\,$GeV), the 
 QCD improved approaches
 \cite{BBNS} used to describe $B$-decays to light mesons 
are not expected to hold. Therefore, we develop a different approach.

As in \cite{EFH} our framework will be threefold: We use the standard
effective Lagrangian approach  for the quark process $b \overline{q} 
\rightarrow c \overline{c}$ (where $q=d,s$) found at a scale below $m_c$
\cite{GKMW}.
Second,  we use  heavy-light chiral perturbation theory for interactions between heavy
mesons and light pseudo-scalar mesons  \cite{HLChPT} to calculate non-factorizable
contributions in terms of chiral loops. Third, to estimate the
contributions from $1/N_c$ suppressed terms at tree level
\cite{EHP,ahjoeB} within heavy-light chiral perturbation theory,
 we use
a recently developed Heavy Light
Chiral Quark Model (HL$\chi$QM) \cite{ahjoe} based on the Heavy Quark
Effective Field Theory (HQEFT) \cite{neu}.

Within our approach, the chiral symmetry breaking scale $\Lambda_\chi
\sim$ 1 GeV is the matching scale for pertubative QCD, chiral
perturbation theory, and HL$\chi$QM. The latter is  also the bridge between 
them.
In the next section 2, we  describe our framework.
In section 3 we present  the factorizable (annihilation), gluon condensate and
chiral loop contributions to the amplitudes. In section 4 we give the
numerical results and conclusions.

\section{Framework}

\subsection{Effective Lagrangian at quark level}

Based on the electroweak and quantum chromodynamical interactions, one
constructs an effective Lagrangian at the quark level in a standard
well known way:
\begin{equation} {\mathcal L}_{W}= \sum_i C_i(\mu) \; Q_i (\mu) \,,
\label{Lquark}
\end{equation}
where $C_i(\mu) = - \frac{G_F}{\sqrt{2}} V_{cb}V_{cq}^* \,
a_i(\mu)$, $q=d,s$ and $a_i(\mu)$ are dimensionless Wilson coefficients that
carry all information of the short distance (SD) physics above the
renormalization scale $\mu$. The matrix elements of  the operators
 $Q_i(\mu)$ on the other hand,
take care of all non-perturbative, long distance (LD) physics below
$\mu$. The relevant operators in our case are 
\begin{eqnarray}
Q_{1}= 4(\overline{q}_L \gamma^\alpha b_L) \; ( \overline{c}_L
\gamma_\alpha c_L )\,,
\qquad   
Q_{2}= 4( \overline{c}_L \gamma^\alpha b_L ) \; ( \overline{q}_L
\gamma_\alpha c_L ) \,, 
\label{Q12} 
\end{eqnarray}  
where  $L$
denotes a left-handed particle. Contributions from  other (say, penguin)
operators are neglected due to smallness of their Wilson coefficients.

In order to obtain all matrix elements of
the Lagrangian (\ref{Lquark}) we need the 
 Fierz transformed version of the operators in  (\ref{Q12}).
To find  these, we use the relation:
\begin{equation}
\delta_{i j}\delta_{l n}  =   \frac{1}{N_c} \delta_{i n} \delta_{l j}
 \; +  \; 2 \; t_{i n}^a \; t_{l j}^a \, ,
\label{fierz}
\end{equation}
where $i$, $j$, $k$ and $n$ are color indices running from 1 to 3 and
$a$ is a color octet index. One obtains
\begin{equation}
Q_1^F = \frac{1}{N_C} Q_2 + 2 \widetilde Q_2 \; ,
  \quad Q_2^F = \frac{1}{N_C} Q_1 + 2 \widetilde Q_1 \; ,
\label{Q12F}
\end{equation}
where the superscript $F$ stands for  ``Fierzed'', and 
\begin{eqnarray}
\widetilde{Q_{1}}  = 4  (\overline{q}_L \gamma^\alpha t^a  b_L )  \; \,
           ( \overline{c}_L \gamma_\alpha t^a c_L ) \,,
\qquad  
\widetilde{Q_{2}}  =  4 \,  ( \overline{c}_L \gamma^\alpha t^a b_L )  \; \,
           ( \overline{q}_L \gamma_\alpha t^a c_L ) \,,
\label{QCol} 
\end{eqnarray}  
where $t^a$ denotes the color matrices.

In order to calculate the matrix elements of the operators
 $\widetilde{Q_{1}}$ and $\widetilde{Q_{2}}$ 
 we will use a version of 
the Heavy Light Chiral Quark Model (HL$\chi$QM)  developed in
\cite{ahjoe}.  It belongs to a class of models extensively
studied in the literature
\cite{chiqm} - \cite{ebert3} and is appropriate 
for describing interactions  {\it in which the transferred enegry is}  of 
the order $1\,$GeV.  This sets the
scale in (\ref{Lquark}) to $\mu \sim
\Lambda_\chi \sim 1\,$GeV, which is by construction the  matching
scale within our approach. At this scale one finds $a_1 \simeq -0.35
-0.07i$ and $a_2 \simeq 1.29+ 0.08i$ \cite{Fleischer,GKMW}.  Note that
the $a_i$'s are complex below the charm scale.

\subsection{Heavy light chiral perturbation theory}

The construction of  heavy light chiral perturbation theory
 is based on the heavy quark
effective theory (HQEFT) \cite{neu}, which is a systematic $1/m_Q$
expansion in the heavy quark mass $m_Q$. The Lagrangian is obtained by 
replacing the heavy quark Dirac field $Q(x)=b(x),c(x)$ or $\overline{c}$ 
with a ``reduced'' field $Q_v^{(+)}(x)$ for a
heavy quark, and $Q_v^{(-)}(x)$ for a heavy anti-quark. These are
related to the full field $Q(x)$ in the following way:
\begin{equation}
Q_v^{(\pm)}(x)=P_{\pm}e^{\mp im_Q v \cdot x} Q(x) \, ,
\end{equation}
where $P_\pm=(1 \pm \gamma \cdot v)/2$ are projecting operators and
$v$ is the velocity of the heavy quark.  
The Lagrangian for heavy quarks then reads:
\begin{equation}
{\mathcal L}_{HQEFT} = \pm \overline{Q_v^{(\pm)}} \, i v \cdot D \,
Q_v^{(\pm)} + {\mathcal O}(m_Q^{- 1}) \; ,
\label{LHQEFT}
\end{equation}
where $D_\mu$ is the covariant derivative containing the gluon field
and ${\mathcal O}(m_Q^{- 1})$ stands for the $1/m_Q$ corrections
\cite{ahjoeB}, which will not be considered in this paper.

After integrating out the heavy and light quarks, the effective Lagrangian
for heavy and light mesons up to ${\mathcal O}(m_Q^{-1})$ can be written as
\cite{ahjoe,itchpt}:
\begin{equation}
{\mathcal L} =  \mp Tr\left[\overline{H^{(\pm)}_{a}}
iv\cdot {\mathcal D}_{ba}
H^{(\pm)}_{b}\right]\, -\, 
\ga \, Tr\left[\overline{H^{(\pm)}_{a}}H^{(\pm)}_{b}
\gamma_\mu\gamma_5 {\mathcal A}^\mu_{ba}\right] \, + .... \, ,
 \label{LS1}
\end{equation}
where $H_a^{(\pm)}$ is the heavy meson field containing a spin zero
and a spin one boson:
\begin{eqnarray}
&H_a^{(\pm)} & \eq P_{\pm} (P_{a \mu}^{(\pm)} \gamma^\mu - i P_{a
5}^{(\pm)} \gamma_5) \; \; .
\label{barH}
\end{eqnarray}
The field $P_M^{(+)}(P_M^{(-)})$ ($M=\mu$ for a vector and $M=5$ for a
pseudo-scalar) annihilates (creates) a heavy meson containing a heavy
quark (anti-quark) with velocity $v$.  Furthermore, ${\mathcal
D}^\mu_{ba}=\delta_{ba} D^\mu-{\mathcal V}^\mu_{ba}$, where $a,b$ are
flavor indices.  The vector and axial vector fields ${\mathcal
V}_{\mu}$ and ${\mathcal A}_{\mu}$ are defined as:
\begin{equation}
{\mathcal V}_{\mu}\eq \fr{i}{2}(\xi^\dagger\partial_\mu\xi
+\xi\partial_\mu\xi^\dagger )\,, \quad {\mathcal A}_\mu\eq -
\fr{i}{2} (\xi^\dagger\partial_\mu\xi -\xi\partial_\mu\xi^\dagger)\,,
 \quad \xi\equiv exp[i\Pi/f] \, ,
\label{defVA}
\end{equation}
where $f$ is the bare pion coupling, and $\Pi$ is a 3 by 3 matrix
which contains the Goldstone bosons $\pi,K,\eta$ in the standard way.
The ellipses in (\ref{LS1}) denote  terms of higher order in the
chiral expansion.

Based on the symmetry of HQEFT, we can obtain the bosonized currents. 
For a decay of the $b \bar{q}$ system we have \cite{ahjoe,itchpt}:
\begin{equation}
 \overline{q_L} \,\gamma^\mu\, Q_{v_b}^{(+)} \; \longrightarrow \;
 \fr{\alpha_H}{2} Tr\left[\xi^{\dagger} \gamma^\alpha L \, H_{b}^{(+)}
 \right] \; ,
\label{J(0)}
\end{equation}
where $\alpha_H=f_H \sqrt{m_H}$.
In the limit $m_Q \rightarrow \infty$, 
$\alpha_B = \alpha_D = \alpha_H$, but there are $1/m_Q$ and
perturbative QCD corrections to this limit \cite{ahjoe,neu}.
Here  $Q_{v_b}^{(+)}$ is the heavy $b$-quark field, $v_b$ is
its velocity, and $H_{b}^{(+)}$ is the corresponding heavy meson
field.

For the $W$-boson materializing to a $\bar{D}$-meson we obtain:
\begin{equation}
 \overline{q_L} \gamma^\mu\  Q_{\bar{v}}^{(-)} \;  \longrightarrow \;
    \fr{\alpha_H}{2} Tr\left[\xi^\dagger \gamma^\alpha L 
 H_{\bar{c}}^{(-)} \right]\,,
\label{Jqc}
\end{equation}
where  $Q_{\bar{v}}^{(-)}$ and $\bar{v} \; ( = v_{\bar c})$ is the
heavy quark field and the   velocity of the  $\bar{c}$ quark,
  respectively,  and 
$H_{\bar{c}}^{(-)}$ is the corresponding field for the  $\bar{D}$-meson.
Similar, for the $b \rightarrow c$ transition, the bosonized current is:
\begin{equation}
 \overline{Q_{v_b}^{(+)}} \,\gamma^\mu\, L Q_{v_c}^{(+)}\;\longrightarrow
 \; - \zeta(\omega) Tr\left[ \overline{H_c^{(+)}}\gamma^\alpha L
 H_{b}^{(+)} \right] \,,
\label{Jbc}
\end{equation}
where  $Q_{v_c}^{(+)}$ and $v_c$ is the heavy quark field
  and the velocity of the
$c$-quark, respectively. Furthermore,
$\zeta(\omega)$ is the Isgur-Wise function for the $\bar{B}
\rightarrow D$  transition,  $H_c^{(+)}$ the heavy $D$-meson
field(s),
and $\omega \equiv v_b \cdot v_c= v_b \cdot
\bar{v}  = 1/(2\kappa)$, where  $\kappa \equiv  M_D/M_B$.  

For the weak current for $D \bar{D}$ production (corresponding to the
factorizable annihilation mechanism) we obtain:
\begin{equation}
 \overline{Q_{v_c}^{(+)}} \,\gamma^\mu\, L Q_{\bar{v}}^{(-)} \; 
 \longrightarrow \;
   - \zeta(-\lambda) Tr\left[
 \overline{H_c^{(+)}} \gamma^\alpha L  H_{\bar{c}}^{(-)} \right]
  \; ,
\label{Jcc}
\end{equation}
where $\lambda= \bar{v}  \cdot v_c = (1/(2\kappa^2)-1)$, and  
$\zeta(-\lambda)$ is a (complex) function less studied and  not so well known
as $\zeta(\omega)$ in (\ref{Jbc}).

\begin{figure}[t]
\begin{center}
\includegraphics[width=9cm]{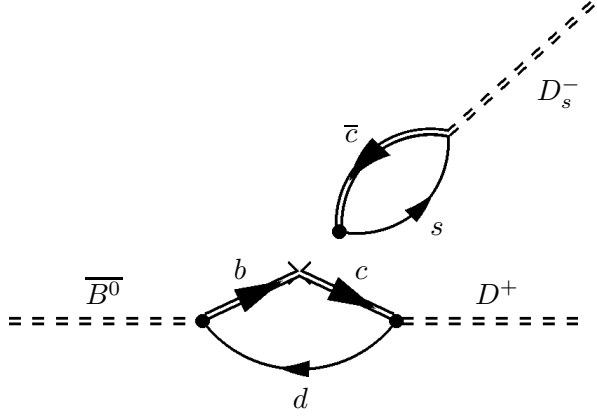}
\caption{\small{Factorized contribution for 
$\overline{B^0}  \rightarrow D^+ D_s^-$
through the spectator mechanism, which does not exist for
 decay mode $\overline{B^0} \rightarrow D_s^+ D_s^-$.}}
\label{fig:bdd_fact}
\end{center}
\end{figure}

\begin{figure}[t]
\begin{center}
\includegraphics[width=8cm]{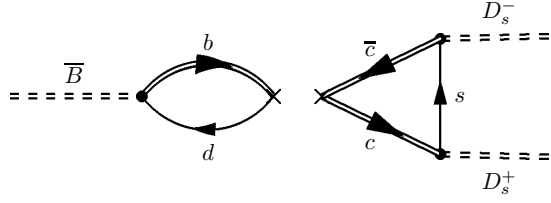} 
\caption{\small{Factorized contribution for 
$\overline{B^0}  \rightarrow D_s^+ D_s^-$
through the annihilation  mechanism, which give zero contributions if
both $D_s^+$ and $D_s^-$ are pseudoscalars.}}
 \label{fig:bdd_fact2}
\end{center}
\end{figure}

The factorized contributions for the spectator and annihilation
diagrams are shown in Figs. \ref{fig:bdd_fact} and \ref{fig:bdd_fact2}.
The first diagram does not give any (direct)
contributions to the class of processes we consider, but is still
important because it is the basis for our chiral loops.

\begin{center}
\begin{figure}
\includegraphics[width=15cm]{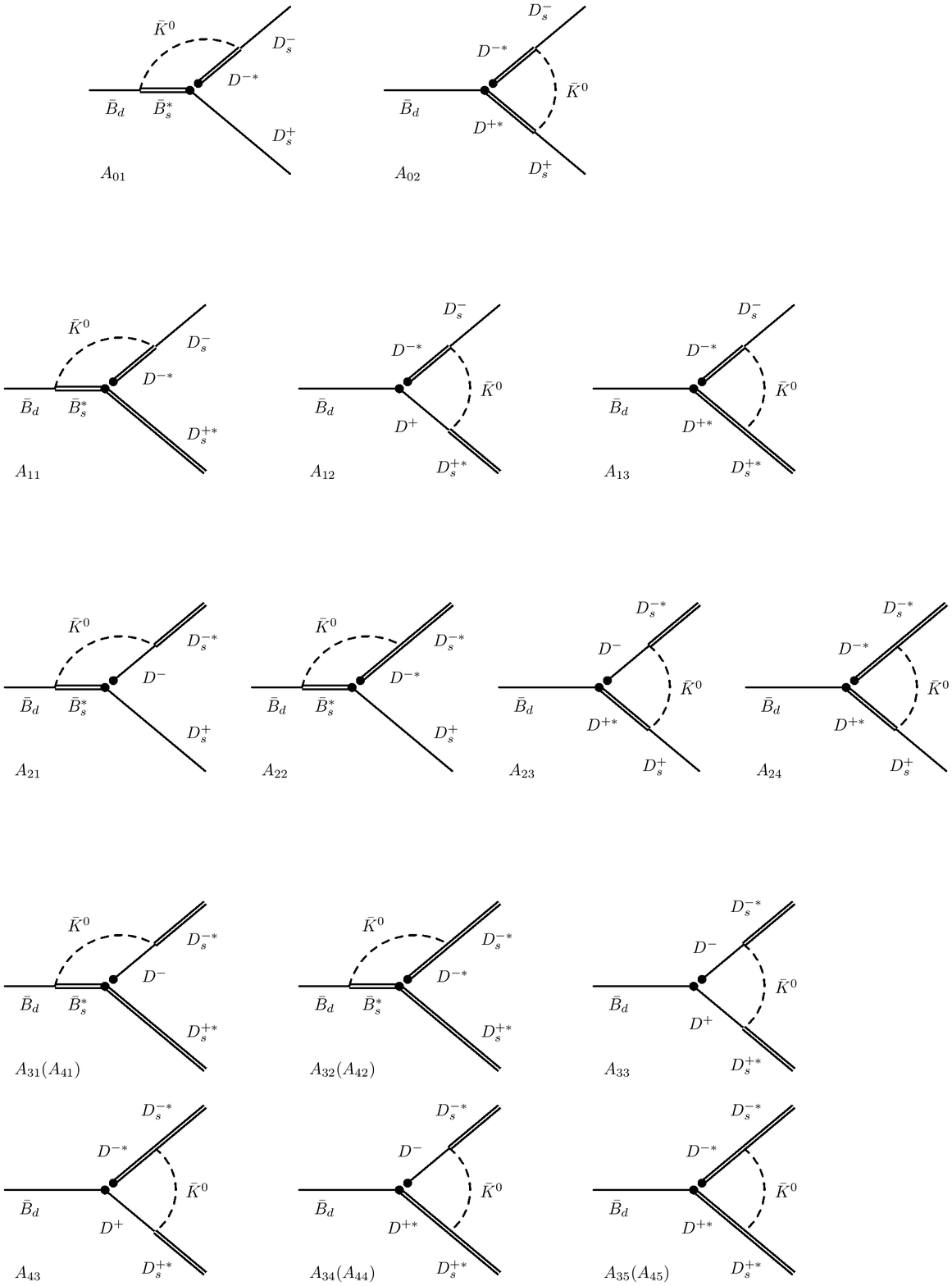}
\caption{Feynman diagrams for the chiral loop contributions.
For $\overline{B^0}  \rightarrow D_s^{+*} D_s^{-*} \,$, the diagrams
 $A_{4i}$ stand for  contributions proportional to the Levi-Civita term.}
\label{grafi}
\end{figure}
\end{center}

The non-factorizable chiral loop contributions to the amplitudes can be visualized in
Fig. \ref{grafi}, where the heavy pseudo-scalar mesons are represented
with single lines, the heavy vector mesons with double lines, the
light kaons with dashed lines, and the weak vertex with two circles.
As seen from Fig.~\ref{grafi},  the chiral loop contribution can be
divided into two topologies that we will denote as topology I and
topology II, respectively. In the topology I (diagrams
$A_{01}$, $A_{11}$, $A_{21}$, $A_{22}$, $A_{31}(A_{41})$, and 
 $A_{32}(A_{42})$ in Fig. \ref{grafi}) the $B$-meson
radiates $\bar K^0$ becoming a $b \bar s$ state that decays into
$D^-(D^{-*})$ and $D_s^+(D_s^{+*})$. The $\bar K^0$ meson is then
reabsorbed by $D^-(D^{-*})$ giving the $D_s^-(D_s^{-*})$ state. In the
topology II (diagrams $A_{02}$, $A_{12}$, $A_{13}$, $A_{23}$,
$A_{24}$, $A_{33}(A_{43}) \,$, $A_{34}(A_{44})$, and 
 $A_{35}(A_{45})$ in Fig. \ref{grafi}),
the  $B$-meson decays into $D^-(D^{-*})$ and $D^+(D^{+*})$
which then re-scatter by means of a kaon into $D_s^-(D_s^{-*})$ and
$D_s^+(D_s^{+*})$.  In both cases, the amplitudes for intermediate
$\bar B^0 \to D^+ D^-$ or $\bar B_s \to D^- D_s^{+}$ processes (and
similar for one or two vectors in a final state) can, within the
factorized limit, be written as the product of currents (\ref{Jqc})
and (\ref{Jbc}) multiplied by the numerically favorable Wilson
coefficient combination $(C_2 + C_1/N_c)$.
Vertices describing absorption or radiation of a 
kaon are given by (\ref{LS1}). There are of course also factorizable
loop contributions, but these are included in the decay constants
$f_{D,B}$ and the Isgur-Wise functions.

The calculation of the chiral loop amplitudes includes the calculation
of a divergent integral of a form:
 \begin{eqnarray}
 I(v_1,v_2)^{\sigma \rho} \; = \; \frac{1}{4}
 \int \frac{d^Dk}{(2 \pi)^D}
 \frac{k^\sigma \, k^\rho}
{(k \cdot v_1 + i \epsilon)(k \cdot v_2 + i \epsilon)(k^2-m_K^2 + i \epsilon)}\,.
\label{loopint1}
\end{eqnarray}
In the dimensional regularization method, the integral can be
rewritten as:
\begin{eqnarray}
\nonumber \\
I(v_1,v_2)^{\sigma \rho} \; = \; \frac{1}{4}
I_1 \left[-r \, g_D^{\sigma \rho} + \frac{r-x}{1-x^2}(v_1^\sigma v_1^\rho
+v_2^\sigma v_2^\rho) + 
\frac{1-x \, r}{1-x^2}(v_1^\sigma v_2^\rho +v_2^\sigma v_1^\rho) \right]\,,
\label{Int}
\end{eqnarray}
where $g_D^{\sigma \rho}$ is a D dimensional matrix tensor and 
\begin{equation}
I_1 \equiv \int \frac{d^Dk}{(2\pi)^D}\frac{1}{k^2-m_K^2}=
\frac{i m_K^2}{16 \pi^2}\left[\Delta-\ln\frac{m_K^2}{\mu^2}+1\right]\,,
\label{I1}
\end{equation}
with $\Delta=2/(4-D)-\gamma_E+\ln 4\pi$. Here $r=r(x)$ with $x=v_1
\cdot v_2$ is a function defined by: 
\begin{equation}
\label{r-def}
r(x)= \frac{1}{\sqrt{x^2-1}} \ln\left(|x|+\sqrt{x^2-1}\,\right)\,,
\label{rf+}
\end{equation}
for $x>0$, and 
\begin{equation}
r(x)= -\frac{1}{\sqrt{x^2-1}} 
(\ln\left(|x|+\sqrt{x^2-1}\,\right)-i\pi)\,,
\label{rf-}
\end{equation}
for $x<0$.

\subsection{The heavy light chiral quark model (HL$\chi$QM)}

The HL$\chi$QM Lagrangian can be written as 
\begin{equation}
{\mathcal L}_{\rm{HL\chi QM}} = {\mathcal L}_{HQEFT} +
{\mathcal L}_{\chi QM} + {\mathcal L}_{Int} \; .
\label{totlag}
\end{equation}
The first term describes the interaction of heavy quarks in
(\ref{LHQEFT}).
The second term describes the interactions of light quarks with light
(Goldstone) mesons in terms of the chiral quark model ($\chi$QM)
\cite{chiqm,bijnes,epb,BEF,pider}:
\begin{equation}
{\mathcal L}_{\chi QM} =
\bar \chi \left[\gamma^\mu (i D_\mu + {\mathcal V}_{\mu}+  
\gamma_5  {\mathcal A}_{\mu}) - m \right]\chi  \; .
\label{chqmR}
\end{equation}
Here $m=0.23\pm0.02\,$GeV is the $SU(3)$ invariant constituent light
quark mass, and $\chi$ is the flavor rotated quark fields given by
$\chi_L=\xi^\dagger q_L$ and $\chi_R = \xi q_R$, where $q^T = (u,d,s)$
are the light quark fields. The left- and right-handed projections
$q_L$ and $q_R$ are transforming after $SU(3)_L$ and $SU(3)_R$
respectively. In (\ref{chqmR}) we have discarded terms involving the
light current quark mass which are irrelevant in the present paper
(but become important when calculating counterterms).
The covariant derivative $D_\mu$ in (\ref{chqmR}) contains the soft
gluon field forming the gluon condensates.
The effects based on (\ref{chqmR})  can be
calculated by Feynman diagram techniques as in
\cite{EHP,ahjoeB,ahjoe,epb,BEF} or by the means
of heat kernel techniques as in \cite{bijnes,ebert3,pider}.

The interaction between heavy meson fields and quarks is
described by \cite{ahjoe,NoZa,barhi,effr,itCQM}: 
\begin{equation}
{\mathcal L}_{Int}  =   
 -   G_H \, \left[ \chibar_a \, \overline{H_a^{(\pm)}} 
\, Q^{(\pm)}_{v} \,
  +     \overline{Q_{v}^{(\pm)}} \, H_a^{(\pm)} \, \chi_a \right] \; ,
\label{L_Int}
\end{equation}
with the coupling constant $G_H = \sqrt{2 m \rho}/f \, $,
where $\rho$ is a hadronic parameter (of the order one) depending on $m$ 
\cite{ahjoe}.

Within the model, one finds the following expression for the Isgur-Wise
function \cite{ahjoe}:
\begin{equation}
\zeta(\omega)= \fr{2}{1+\omega} \left(1-\rho \right)+
 \rho \, r(\omega)\, ,
\label{iw}
\end{equation}
where $r(\omega)$ is given by (\ref{rf+}). 
In the  simple expression  (\ref{iw}) 
chiral loop and perturbative QCD corrections down to the scale 
$\mu= \Lambda_\chi$, have  not been taken into account.
 Nevertheless, it
gives a good description of the Isgur-Wise function
\cite{ahjoe}. For negative values of the argument 
the Isgur-Wise function $\zeta(-\lambda)$ is,
  within the HL$\chi$QM \cite{ahjoe},  still given by (\ref{iw}) with 
$\omega \rightarrow  -\lambda$.

Performing the bosonization of the HL$\chi$QM, one encounters
divergent loop integrals which will in general be quadraticly, linearly
and logarithmicly divergent \cite{ahjoe}. The quadraticly and
logarithmicly divergent integrals are related to the
quark condensate and  the bare pion decay constant $f$ \cite{BEF},
respectively.
The linearly  divergent integral (which is finite within
dimensional
regularization) is related to the axial coupling $\ga$ in (\ref{LS1}).

\begin{figure}[t]
\begin{center}
\includegraphics[width=9cm]{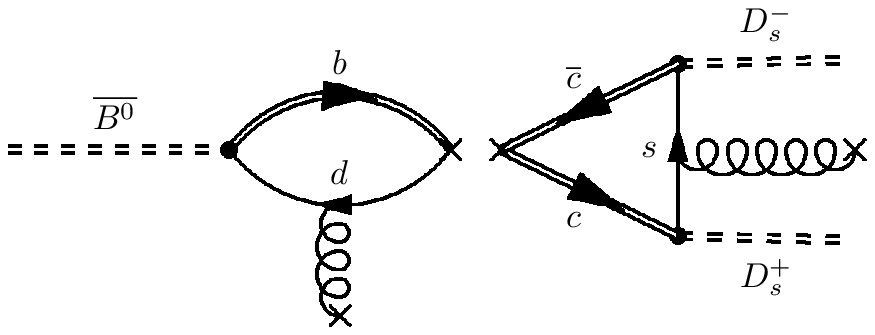}
\caption{Non-factorizable contribution for 
$\overline{B^0}  \rightarrow D_s^+ D_s^-$
through the annihilation mechanism with additional soft gluon emision.
 The wavy lines represent soft
gluons ending in vacuum to make gluon condensates.}
\label{fig:bdd_nfact2}
\end{center}
\end{figure}

The gluon condensate contribution to the amplitudes can be written, 
within the
framework presented in the previous section, in a quasi-factorized
way as a product of matrix elements of colored currents:
\begin{eqnarray}
\langle D_s^+ D_s^- | {\mathcal L}_W| \overline{B^0} \rangle_{NF}^G \, = \,
  8 C_2 \,
\langle D_s^+ D_s^-|\overline{c}_L \gamma_\mu t^a c_L |G \rangle
 \langle G |\overline{d}_L \gamma^\mu t^a b_L |\overline{B^0} \rangle
\; ,
\label{QGlue}
\end{eqnarray}
where $G$ in the bra-kets symbolizes emission of a gluon as visualized
in Fig. \ref{fig:bdd_nfact2}.  The left part in
Fig. \ref{fig:bdd_nfact2} gives us the bosonized colored current:
\bea
&&\left(\overline{q_L}\, t^a  \,\gamma^\alpha \, Q_{v_b}^{(+)}\right)_{1G} 
\;   \longrightarrow \; 
- \fr{G_H \, g_s}{64 \pi} \,G_{\mu\nu}^a Tr\left[\xi^\dagger
\gamma^\alpha L \, H_b^{(+)}
\left( \sigma^{\mu\nu} \, - \, F \,  \{\sigma^{\mu\nu},
 \gamma \cdot v_b \} \, \right)\right] \,,
\label{1G}
\eea
where $G^a_{\mu \nu}$ is the octet gluon tensor, and
$F \; \equiv \; 2 \pi f^2/(m^2\,N_c)$
is a dimensionless quantity of the order 1/3. The symbol $\{\; , \; \}$
denotes the anti-commutator.


For the creation of a $D \bar{D}$ pair in the right part of Fig.
\ref{fig:bdd_nfact2} (the analogue of (\ref{1G})) one gets:
\bea
 \left(\overline{Q_{v_c}^{(+)}} \,t^a \; 
\gamma^\alpha \, L Q_{\bar{v}}^{(-)}\right)_{1G} \;
\;   \longrightarrow \;  
 \fr{G_H^2 \, g_s}{128 \pi m (\lambda-1)}\,G_{\mu\nu}^a
\;Tr\left[ \overline{H_c^{(+)}} \; \gamma^\alpha \, L \,  
  H_{\bar{c}}^{(-)}  \right.  \nonumber \\
\left.  \times  \left( X \sigma^{\mu\nu}
\, + \,
 \left\{\sigma^{\mu\nu}, \gamma \cdot \Delta v \right\}  \,
  \right)\right] \; ,
\label{1Gc}
\eea
where 
\begin{equation}
X \equiv \frac{4}{\pi} (\lambda -1) \, r(-\lambda) \; ,
\label{X}
\end{equation}
and $\,\Delta v= v_c-\bar{v} \,$.
Multiplying the currents in Eqs. (\ref{1G}) and (\ref{1Gc}),
and using   the replacement:
\begin{equation}
g_s^2 G_{\mu \nu}^a G_{\alpha \beta}^a  \; \rightarrow 4 \pi^2
 \gc \frac{1}{12} (g_{\mu \alpha} g_{\nu \beta} -  
g_{\mu \beta} g_{\nu \alpha} ) \, ,
\label{gluecond}
\end{equation}
we obtain a bosonized effective Lagrangian term which is $1/N_c$
suppressed compared to the factorized contributions. 
This effective Lagrangian term corresponds to a certain linear combination 
of  a priori possible  $1/N_c$ suppressed terms  at the tree level in
the chiral perturbation theory sense. 

\section{Calculating the decay amplitudes}

First, for comparison,  we give the  factorized amplitude
for the process $\overline{B^0} \rightarrow D^+ D_s^-$ 
\bea
{\cal A}(\overline{B^0} \rightarrow D^+ D_s^-)_F = 
 (C_2 + \frac{C_1}{N_c})\,\zeta(\omega) f_D M_D 
\sqrt{M_B M_D} \; (\lambda + \omega) \; \, .
\label{BDDF}
\eea
\noindent
The $\bar B^0 \to M \bar M$, $M=D_s^+, D_s^{+*}$ decay amplitudes can
be written in the following form:
\begin{equation}
{\cal A}(B \to D \bar D)=F^{PP}_0, \qquad
{\cal A}(B \to D^* \bar D)= i F^{VP}_0 \;\varepsilon_{D^*} \cdot \bar{v}\,, 
\end{equation} 
\begin{equation}
{\cal A}(B \to D \bar{D^*})= i F^{VP}_0 \;\varepsilon_{\bar{D^*}} \cdot v_{c}\,, 
\end{equation} 
\begin{equation}
{\cal A}(B \to D^* \bar{D^*})=F^{VV}_0 \;\varepsilon_{D^*} \cdot
\varepsilon_{\bar{D^*}}+F^{VV}_1 \;\varepsilon_{D^*} \cdot \bar{v}
\;\varepsilon_{\bar{D^*}} \cdot v_{c} + F^{VV}_2 \;i \epsilon_{\alpha\beta\mu\nu}
v_c^\alpha \bar{v}^\beta \varepsilon_{D^*}^\mu  \varepsilon_{\bar{D^*}}^\nu\,,
\end{equation} 
with the reduced amplitudes $F^{NN}_i$,
containing the contributions 
coming from the tree level factorizable contributions $F^{NN}_{i,fc}$, 
gluon condensates $F^{NN}_{i,gc}$  and chiral loops $F^{NN}_{i,cl}$:
\begin{equation} 
F^{NN}_i=F^{NN}_{i,fc}+F^{NN}_{i,gc}+F^{NN}_{i,cl} \; ,
 \end{equation} 
 where $N=P,V$  mean
  pseudoscalar $D$ and vector meson $D^*$, respectively.

The factorizable  amplitude for  $\overline{B^0} \to M \bar M$, $M=D^+_s,D^{+*}_s$ 
comes from the annihilation diagram:
 \begin{eqnarray}
\langle D_s^- D_s^+ | {\mathcal L}_W| \overline{B^0} \rangle_F \, = \, 
 4 (C_1 +\frac{1}{N_c} C_2) 
\langle D_s^- D_s^+|\overline{c_L} \gamma_\mu  c_L |0 \rangle
 \langle 0 |\overline{d_L} \gamma^\mu  b_L |\overline{B^0} \rangle \;, 
\label{FactorizAnn}
\end{eqnarray}
and is proportional to a  numerically non-favorable combination of Wilson coefficients.
Using Eqs. (\ref{J(0)}) and (\ref{Jcc}) we obtain the
following values for the reduced amplitudes:
\begin{equation}
F^{PV}_{0,fc}=-F^{VP}_{0,fc}=F^{VV}_{2,fc}\sqrt{\frac{M_D}{M_{D^*}}} \; = \;   
2 \,(C_1 + \frac{C_2}{N_c})\,\zeta(-\lambda)\alpha_B\,\kappa\; 
 \,  \sqrt{M_B M_D M_{D^*}}\,,
\end{equation}
\begin{equation}
F^{PP}_{0,fc}=F^{VV}_{0,fc}=F^{VV}_{1,fc}=0\, .
\end{equation}

Using   (25) - (28), we obtain the
gluon condensate contributions (the color suppressed $1/N_c$
contributions
at the  tree level)  
to the reduced amplitudes
illustrated in Fig. \ref{fig:bdd_nfact2}:
\begin{eqnarray}
F^{PP}_{0,gc} \; = \; 3 \, S \, \left(X+
 \frac{4}{3} (\lambda-1) \right) \,, 
\qquad
F^{VP}_{0,gc} \; = \, S \, \left(X \left[ 1-2 F \right] 
+ 4 \left[ \lambda + 2 F \right] \right)\,,
\label{BDDG0}
\end{eqnarray}
\begin{eqnarray}
 F^{PV}_{0,gc} \; = \; S \,
 \left(X \left[ 3+2 F \right] + 4 \left[ (\lambda-2) - 2 F \right]
 \right) \,, \qquad
F^{VV}_{0,gc} \; = \; 2 \kappa^2 S \, X \, (\lambda+1) \,,
 \label{BDDG2}
\end{eqnarray}
\begin{eqnarray}
F^{VV}_{1,gc} \; = \; -2 \kappa^2 S \, (X-4)\,,
\qquad
F^{VV}_{2,gc} \; = \; 2  \kappa^2 S \, (X-4)(1+2 F)\,,
 \label{BDDG3}
\end{eqnarray}
where $F$ is defined below (\ref{1G}),  $X$ is defined in (\ref{X}), and
\begin{equation} 
 S \, \equiv \, \frac{C_2 \, \left(G_H \sqrt{M_B}\right)^3}{3 \cdot
 2^9 m (\lambda-1)} \; \gc\,. 
\label{Sdef} 
\end{equation}

In the evaluation of the integrals (\ref{Int}) and (\ref{I1})
appearing in the reduced chiral loops amplitudes $F^{NN}_{i,cl}$, we
use the  $\overline{MS}$ scheme and we take $\mu \simeq \Lambda_{\chi}
\simeq 1\,$ GeV.  The integral (\ref{I1}) contains a logarithmicly
divergent term and a constant term. However, additional
contributions to the constant term might come from counterterms. The
finite part of these terms is unknown and therefore, in our numerical
computation, we take into account logarithmic terms only, which are
independent of the counterterms contributions, and we consider
constant term contributions as theoretical uncertainities of our
approach.

However, the products of
two Levi-Civita terms enter  in our computation of the amplitudes coming from diagrams
$A_{13}\,,A_{22}\,,A_{32}\,,A_{35}$ and $A_{45}$. Whitin dimensional
regularization, these
products are not uniquelly defined. This is related to the problem
known in the literature as $\gamma_5$ scheme dependance
\cite{life}. However,  this scheme dependence appears in the
constant terms only. To estimate its influence on numerical results,
we use two different appraches to the products of two Levi-Civita
symbols. (One is the dimensoinal reduction \cite{life} and the 
second one is a variation of the dimensional reduction in which the products of 
metric tensors in D dimension and the 
metric tensor in 4-dimensions are fixed on the 4-dimensional space.) 

The logarithmic contributions to the reduced
amplitudes are:

\begin{eqnarray}
 F^{PP}_{0,cl}\; &=& \;   
iK \,\sqrt{\frac{M_D}{M_D^*}} \,\left[-2(1+\omega) \left(r(-\omega)+r(-\lambda)\right)
+2 (\omega+\lambda) \right],
 \label{AmpBDDa} \\
  F^{VP}_{0,cl}\; &=& \;  -i 
K\, \left[-2 \kappa \left(r(-\omega)+r(-\lambda)\right)
+2 (\kappa + 1) \right],
 \label{AmpBDDb} \\
 F^{PV}_{0,cl}\; &=& \; -i  
K\,\left[2 \kappa \left(r(-\omega)+r(-\lambda)\right)
+2 (\kappa + 1) \right],
 \label{AmpBDDc} \\
F^{VV}_{0,cl}\; &=& \;  
iK\,\sqrt{\frac{M_D^*}{M_D}} \, \left(2(\omega+1)\right), \\
F^{VV}_{1,cl}\; &=& \; 
iK\,\sqrt{\frac{M_D^*}{M_D}}
\left((\kappa+1)2\left[r(-\omega)+r(-\lambda)\right]
-2 (\omega + \lambda) (H+G) -2\kappa^2 \right), \\
F^{VV}_{2,cl}\; &=& \; 
-i K \,\sqrt{\frac{M_D^*}{M_D}} 2 \kappa, 
 \label{AmpBDDd}
\end{eqnarray}
where
\begin{eqnarray}
K \equiv \frac{1}{2}C_2 \, \zeta(\omega) \alpha_D 
\left(\frac{g_A}{f}\right)^2 I_L \, \sqrt{M_B M_{D^*} M_D} \, , \qquad
I_L \equiv \frac{-i m_K^2}{16 \pi^2} \ln\frac{m_K^2}{\mu^2}\, ,
\label{KI}
\end{eqnarray}
\begin{eqnarray}
G  \equiv \frac{(r(-\omega)+\omega)}{1-\omega^2} \, \kappa^2  -
\frac{(1+\omega \, r(-\omega))}{1-\omega^2} \,\kappa
\; \; , \quad
H \equiv - \frac{(1+\lambda \, r(-\lambda))}{1-\lambda^2}\,.
\label{GH}
\end{eqnarray}
Note that $K$ is a priori proportional to the Wilson coefficient
combination $(C_2 + C_1/N_c) \,$,  as the amplitude in (\ref{BDDF}). But
because the factor $1/f^2$ is already of order $1/N_c \, $, we replace 
 $(C_2 + C_1/N_c)$ by just $C_2$ in  (\ref{KI}).

\section{Results and discussion}

In our calculation we used the following input parameters: 
$\alpha_B=\alpha_D=0.33$ GeV$^{-3/2}$, $\rho=1.05$, $G_H=7.5$ GeV$^{-1/2}$ 
and $\gc= [(0.315\pm0.020)$ GeV]$^{4}$
\cite{ahjoeB,ahjoe}, $\ga=0.6$ \cite{ga}, $f_\pi=0.093$ GeV  \cite{PDG}
and $\kappa=0.37$. 
\begin{table}
\begin{center}
\begin{tabular}{|c|c|c|c|c|c|c|}
\hline
$i$ & $F^{PP}_{0,i} \times 10^7$ & $F^{VP}_{0,i} \times 10^7$ & $F^{PV}_{0,i}\times 10^7$ & 
$F^{VV}_{0,i}\times 10^7$& $F^{VV}_{1,i}\times 10^7$ & $F^{VV}_{2,i}\times 10^7$ \\ \hline
$fc$ & 0 & 0.12i  & -0.12i & 0 &  0 & -0.12i \\
$gc$ & -0.17-0.71i & 1.11 + 0.14i & -0.51 + 0.87i & 0.13-0.22i &
-0.13+0.06i & 0.22 -0.09i \\
$cl$ & 0.91-1.20i &-0.23 + 0.19i & -0.08 - 0.21i & 0.56+0.03i &  -0.24-0.45i & -0.09 \\  
\hline
\end{tabular}
\caption{Reduced amplitudes for the $B^0 \to M \bar M$, $M=D_s^+, D_s^{+*}$
decay modes. The results are given in GeV.}
\end{center}
\end{table}

\begin{table}
\begin{center}
\begin{tabular}{|c|c|c|c|c|c|c|}
\hline
$i$ & $F^{PP}_{0,i} \times 10^7$ & $F^{VP}_{0,i} \times 10^7$ & $F^{PV}_{0,i}\times 10^7$ & 
$F^{VV}_{0,i}\times 10^7$& $F^{VV}_{1,i}\times 10^7$ & $F^{VV}_{2,i}\times 10^7$ \\ \hline
$fc$ & 0 & 0.51i & -0.51i & 0 &  0 & -0.51i \\
$gc$ & -0.74-3.16i & 4.95 +0.63i & -2.27 +3.69i & 0.60-1.00i &
-0.59+0.25i & 0.99 -0.42i \\
$cl$ & 3.78-4.96i & -0.96 +0.78i & -0.34 - 0.85i & 2.13+0.10i &
-0.91-1.72i & -0.34 -0.02i \\
\hline
\end{tabular}
\caption{Reduced amplitudes for the $B_s^0 \to N \bar N$, $N=D^+, D^{+*}$
decay modes. The results are given in GeV.}
\end{center}
\end{table}
The reduced amplitudes for $B^0 \to M \bar M$, $M=D_s^+, D_s^{+*}$ are
presented in Table 1. 
 We find the following branching ratios:
\begin{eqnarray}
Br(\bar B^0 \to D_s^+ D_s^-)=2.5 \times 10^{-4}\,, \qquad Br(\bar B^0_s \to D^+ D^-)=4.5 \times 10^{-3}\,,\\
Br(\bar B^0 \to D_s^{+*} D_s^-)=3.3 \times 10^{-4}\,,\qquad  Br(\bar B^0_s \to D^{+*} D^-)=6.8 \times 10^{-3}\,,\\
Br(\bar B^0 \to D_s^+ D_s^{-*})=2.0 \times 10^{-4}\,,\qquad Br(\bar B^0_s \to D^+ D^{-*})=4.3 \times 10^{-3}\,, \\
Br(\bar B^0 \to D_s^{*+} D_s^{-*})=5.4\times 10^{-4}\,,\qquad Br(\bar B^0_s \to D^{*+} D^{-*})=9.1 \times 10^{-3}\,.
\end{eqnarray}

The contribution of the constant term and the corresponding counterterm
can change the branchig ratio for $B$-meson decaying into two pseudoscalars by about
$10\%$, while in the case of decay  into one pseudoscalar and one
vector $D$-meson, this contribution is  in the range of  
$20-40\%$. 
In the case of $B$-meson decaying  into two vector mesons, the 
constant term is
estimated to be 2-8 times larger than the logaritmic contribution,
depending on  the  choice of the scheme in which the products of two Levi-Civita 
terms 
are considered.
The uncertainty in input parameters can result in an additional error 
 for the branching ratios. We estimate that it can be of the order of 
$~ 20\%$. 
Within our approach the 
 $1/m_Q$ corrections, with  $Q=c,b$   have  been  neglected. 
At least the $1/m_c$ corrections might be important.

The study of  dominant contributions in the 
 $\bar B^0_d \to D_s^+ D_s^-$, 
  $\bar B^0_d \to D_s^{+*} D_s^- $ and 
 $\bar B^0_d \to D_s^+ D_s^{-*}$ 
decay amplitudes  is very important for our understaning of the color suppressed 
 contributions to the $B$-mesons decaying to two charm mesons. 
This is even more important knowing that the experimental rates for $B_d$
decay into $D^- D_s^+$, $D^{-*} D_s^+$, $D^- D_s^{*+}$ and $D^{*-}
D_s^{*+}$ have very small color suppressed 
contributions and therefore the decay amplitudes we consider 
open a window for
 studies of  color suppressed effects in $B$-decays to two charm mesons.  
The chance for experimental measurements of these decay rates at $B$-factories 
makes the study of the decay mechanisms in these decays even more important. 

\section{Appendix A: Chiral loops with ``superpropagator''}

In the following, we will shortly present a
 ``superpropagator method'' which enables us to calculate all
contributions of one kind of topology to all decays of a type $\bar
B^0 \to M \bar M$, $M= D_s^*,D_s^{+*}$ as one compact calculation. This means that
instead of calculating all the diagrams in Fig. \ref{grafi}, we only
need to calculate two contributions, each one coming from one
topology. 

The superpropagator is a propagator of the $H^{(\pm)}(x)$ field and
therefore includes propagators of both heavy pseudo-scalar and heavy
vector meson. It is defined in a standard way as a contraction of
$H^{(\pm)}(x)_{\alpha \beta}$ and $\overline{H^{(\pm)}}(y)_{\kappa
\lambda}$, where $\alpha, \beta, \kappa, \lambda$ are Dirac spinor
indices. In momentum space, we obtain the superpropagator
 \begin{equation}
 S^{(\pm)}_{\alpha \beta ; \kappa \lambda}(k) \; 
= \; \frac{1}{2(\pm v \cdot k + i \epsilon)} \;  
 T^{(\pm)}_{\alpha \beta ; \kappa \lambda}(v)\,,
 \label{prop}
\end{equation}
where $k$ is the momentum and 
 \begin{eqnarray}
 T^{(+)}_{\alpha \beta ; \kappa \lambda}(v) \; 
=  \;  
 \left( P_{+}(v) \gamma_\tau \right)_{\alpha \beta}
(-g^{\tau \nu}+v^\tau v^\nu) 
\left( \gamma_\nu P_{+}(v)  \right)_{\kappa \lambda}
\nonumber \\
- \left(  P_{+}(v) \gamma_5 \right)_{\alpha \beta} \; 
\left( \gamma_5 P_{+}(v) \right)_{\kappa \lambda}\,,
\label{Tplus}
\end{eqnarray}
 \begin{eqnarray}
 T^{(-)}_{\alpha \beta ; \kappa \lambda}(v) \; 
=  \;  
 \left(  \gamma_\tau P_{-}(v) \right)_{\alpha \beta}
(-g^{\tau \nu}+v^\tau v^\nu) 
\left(  P_{-}(v) \gamma_\nu \right)_{\kappa \lambda}
\nonumber \\
- \left( \gamma_5 P_{-}(v)  \right)_{\alpha \beta} \; 
\left( P_{-}(v)  \gamma_5 \right)_{\kappa \lambda}\,.
\label{Tmin}
\end{eqnarray}
For the effective Lagrangian of topology I we have:
\begin{eqnarray}
 {\mathcal L}_I(B \rightarrow D \bar{D}) \; &=& \;   
 -i \, C_2 \, \zeta(\omega)\fr{\alpha_D}{2} (\frac{g_A}{f})^2 \;
  I(v_b,-\overline{v})^{\sigma \rho}
 \left[\overline{H_c^{(+)}} \gamma^\mu L \right]_{ \beta \alpha}
 T^{(+)}_{\alpha \beta ; \kappa \lambda}(v_b) \nonumber \\
 &\times& \left[H_{b}^{(+)} \gamma_\sigma \gamma_5 \right]_{\lambda \kappa}
\left[H_{\bar{c}}^{(-)} \gamma_\rho \gamma_5 \right]_{\eta \delta} 
 T^{(-)}_{\delta \eta ; \phi \xi}(\overline{v}) \; 
 \left[ \gamma_\mu L \right]_{\xi \phi} \; \; , 
 \label{BDD1}
\end{eqnarray}
while the effective Lagrangian of topology II can be written as:
\begin{eqnarray}
 {\mathcal L}_{II}(B \rightarrow D \bar{D}) \; &=& \;   
 -i \, C_2 \, \zeta(\omega)\fr{\alpha_D}{2} (\frac{g_A}{f})^2 \;
  I(v_c,-\overline{v})^{\sigma \rho}
 \left[\gamma_\sigma \gamma_5 \overline{H_c^{(+)}}\right]_{\beta \alpha}
 T^{(+)}_{\alpha \beta ; \kappa \lambda}(v_c) \nonumber \\
 &\times& \left[\gamma_\mu L H_{b}^{(+)} \right]_{\lambda \kappa}
 \left[H_{\bar{c}}^{(-)} \gamma_\rho \gamma_5 \right]_{\eta \delta}
  T^{(-)}_{\delta \eta ; \phi \xi}(\overline{v}) \; 
\left[ \gamma_\mu L \right]_{\xi \phi}\,.
 \label{BDD2}
\end{eqnarray}
From (\ref{BDD1}) and (\ref{BDD2}) one can derive the (reduced) amplitudes
already given in (\ref{AmpBDDa})-(\ref{AmpBDDd}).

\end{document}